\documentclass[pdflatex,sn-mathphys]{sn-jnl}

\usepackage{amsfonts}
\usepackage{subcaption}
\usepackage{physics}

\newcommand{\PT}{\mathcal{PT}}
\newcommand{\Psym}{\mathcal{P}}
\newcommand{\Tsym}{\mathcal{T}}

\begin{document}

\title{Exceptional point rings and $\PT$-symmetry in the non-Hermitian XY model}


\author{\fnm{Robert A.} \sur{Henry}}

\author{\fnm{D. C.} \sur{Liu}}



\author*{\fnm{Murray T.} 
\sur{Batchelor}}\email{murray.batchelor@anu.edu.au}

\affil[]{\orgdiv{Mathematical Sciences Institute}, \orgname{Australian National University}, \orgaddress{
		\city{Canberra}, 
		\state{ACT 2601}, \country{Australia}}}


\abstract{The XY spin chain is a paradigmatic example of a model solved by free fermions, in which the energy eigenspectrum is built from combinations of quasi-energies. In this article we show that by extending the XY model's anisotropy parameter $\lambda$ to complex values, it is possible for two of the quasi-energies to become degenerate. In the non-Hermitian XY model these quasi-energy degeneracies give rise to exceptional points (EPs) where two of the eigenvalues and their corresponding eigenvectors coalesce. The distinct $\lambda$ values at which EPs appear form concentric rings in the complex plane which are shown in the infinite system size limit to converge to the unit circle coinciding with the boundary between distinct topological phases. The non-Hermitian model is also seen to possess a line of broken $\PT$ symmetry along the pure imaginary $\lambda$-axis. For finite systems, there are four EP values on this broken $\PT$-symmetric line if the system size is a multiple of 4.}




\maketitle

\section{Introduction}

In recent years there has been extensive activity around non-Hermitian models including both theoretical and experimental developments \cite{ashidaNonHermitianPhysics2020, bergholtzExceptionalTopologyNonHermitian2021}. Particularly interesting features involve exceptional points (EPs), which are isolated points (or manifolds) in a model's parameter space where the model Hamiltonian's Jordan block structure becomes non-trivial, i.e., when two of the eigenvalues and their corresponding eigenvectors coalesce~\cite{Heiss2012,bergholtzExceptionalTopologyNonHermitian2021,Ma2022,znojil2020,znojil2022,Sun2024,Feng2025}. This is impossible in Hermitian systems. EPs have many interesting physical properties such as geometric phases and enhanced sensitivity, and can be observed in various experiments (see, e.g., Refs~\cite{Hahn2016,Nag2019,Liang2023,Wang2023,Even2024,Wu2025}.
On the theoretical side, 
these aspects have been mostly studied in relatively simple non-Hermitian few-body systems. Most recently, for example, EPs have been studied~\cite{Sirker2024} in a dimerized Hatano-Nelson model~\cite{HN1996}, or equivalently the Su-Schrieffer-Heeger model~\cite{SSH1979} with asymmetric hopping. EPs have also been studied in other variants of the SSH model (see, e.g.,~\cite{McCann2025} and references therein).
Such investigations can be extended to quantum many-body systems. Recent work in this direction~\cite{henry2023exceptional} has shown that the non-Hermitian extension of the $Z(N)$ Baxter-Fendley free parafermion model~\cite{Baxter1989,Fendley2014,Brief_History} has a ring of exceptional points in the complex plane, including in the $N=2$ case of the non-Hermitian quantum Ising chain.

The non-Hermitian extension of a given model involves extending key model parameters to complex values, thereby producing a complex eigenspectrum in contrast to the real eigenspectrum characteristic of the more well studied Hermitian systems. The inherent structure of the non-Hermitian free parafermion energy eigenspectrum, and the free fermion structure in the Ising case, is particularly convenient for establishing the location of EPs. 
It was shown that when two quasi-energies become degenerate, such  
quasi-energy EPs give rise to Hamiltonian EPs~\cite{henry2023exceptional}.

Here we examine the location of EPs in the energy eigenspectrum of another paradigmatic many-body system -- the anisotropic Heisenberg XY spin chain -- solved long ago by Lieb, Schultz and Mattis~\cite{Lieb1961} by mapping to \textit{free fermions} via the Jordan-Wigner transformation~\cite{Jordan1928} and Bogoliubov transformation in Fourier space. 
As we shall see, the non-Hermitian extension of the anisotropic XY model exhibits two concentric rings of complex $\lambda$ values at which EPs appear, which each approach the unit circle in the infinite size limit. 

An important and extensively investigated aspect of non-Hermitian physics involves $\PT$-symmetric non-Hermitian systems~\cite{benderRealSpectraNonHermitian1998, benderIntroductionSymmetricQuantum2005,El-Ganainy2018}, in which parity-time symmetry can be used as a replacement for Hermiticity to define unitary time evolution, provided a different $\PT$-constructed inner product is used. 
We recall that if a $\PT$-symmetric Hamiltonian is in the unbroken $\PT$ symmetry regime, its eigenvalues are entirely real. 
On the other hand, if the $\PT$ symmetry is broken, the spectrum can be complex, with any non-real eigenvalues appearing in complex conjugate pairs. 
Here we establish that the XY model enjoys broken $\PT$ symmetry  when the complex parameter $\lambda$ is pure imaginary. 
We find that this line of broken $\PT$ symmetry in the complex plane includes a subset of EPs.

\section{The non-Hermitian XY model}

The anisotropic XY model is defined in terms of the Pauli matrices by the Hamiltonian
\begin{equation}
	\label{eq:1-1}
	H_\gamma = -\dfrac{1}{2}\sum_{j=1}^{L-1}\left(\dfrac{1+\gamma}{2}\sigma_j^x\sigma_{j+1}^x + \dfrac{1-\gamma}{2}\sigma_j^y\sigma_{j+1}^y  \right) \,.
\end{equation}
For convenience we assume the chain size $L$ is even.
In this study we employ open boundary conditions, for which the underlying free fermion structure is readily apparent for finite-size systems. 
As mentioned above, Lieb, Schultz and Mattis~\cite{Lieb1961}~showed that the energy eigenspectrum of Hamiltonian (\ref{eq:1-1}) can be expressed in terms of {free fermions}. Key elements of their solution for open boundary conditions are summarised in this section. We rescale the Hamiltonian as follows:
\begin{equation}
	\label{eq:1}
	H_\lambda = -\sum_{j=1}^{L-1}\left(\sigma_j^x\sigma_{j+1}^x + \lambda \, \sigma_j^y\sigma_{j+1}^y  \right),
\end{equation}
where it is convenient for the purposes of this work to define
\begin{equation} \label{trans}
\lambda = \frac{1-\gamma}{1+\gamma} \quad \mathrm{with} \quad H_\lambda = \frac{4}{1+\gamma} H_\gamma \,. 
\end{equation}
Lieb et al.~assumed real values of $\gamma$, but their results extend directly to complex $\gamma$ or $\lambda$, which allows the calculation of the model's EPs as shown in Section~\ref{sec:eps}. 

Various non-Hermitian extensions of the XY model have been considered previously. 
Hamiltonian (\ref{eq:1-1}) was studied for pure imaginary $\gamma$ with a real transverse magnetic field subject to periodic boundary conditions~\cite{Zhang2013,Miao2024}. 
Other non-Hermitian extensions of the more general XYh model have also been considered (\cite{Liu2025} and references therein). 

\subsection{Free fermion structure}

In any finite size free fermion system with open boundary conditions, each Hamiltonian energy eigenvalue is precisely of the form
\begin{equation}
	\label{eq:ff}
	E = \sum_{j=1}^L \pm \, \epsilon_j,
\end{equation}
giving $2^L$ combinations in total. 
In this way each \textit{quasi-energy} $\epsilon_j$ is a 
fundamental building block making up the energy eigenspectrum. 
The groundstate energy is the particular combination
\begin{equation}
  \label{eq:2}
  E_0 = -\sum_{j=1}^L \epsilon_j.
\end{equation}

For Hamiltonian (\ref{eq:1}) the quasi-energies are given by $\epsilon_j = a_j^{1/2}$, where $a_j$ is an eigenvalue of $C^TC$ with $C$ the $L\times L$ matrix
\begin{equation}
	\label{eq:9}
	\mathcal{C} = 
    \begin{pmatrix}
		0        & 1 & {}       & {}       & {}       & {}         \\
		\lambda & 0        & 1 & {}       & {}       & {}         \\
		{}       & \lambda & 0        & 1 & {}       & {}         \\
		{}       & {}       & \ddots   & \ddots   & \ddots   & {}         \\
		{}       & {}       & {}       & \lambda & 0        & 1 & \\
		{}       & {}       & {}       & {}       & \lambda & 0
	\end{pmatrix} \,.
\end{equation}
Diagonalising the $L \times L$ quasi-energy matrix is an efficient way to examine the model's eigenvalues numerically, and this is how numerical data is produced in this article   unless stated otherwise. Because of the staggered form of $\mathcal{C}$, its eigenvectors form two separate sectors.

\subsection{Quasi-momenta}

A key role in the exact solution of the XY model is played by the quasi-momentum $k$, which is used to determine the eigenspectrum via the quasi-energies. 
Specifically, for the open boundary conditions under consideration, the quasi-momenta $k_j$ are the $L$ solutions of the equation
\begin{equation}
	\label{eq:k_cond}
	\dfrac{\sin{(L+2)k}}{\sin{L k}} = -  \lambda^{\pm1},
\end{equation}
where $+$ and $-$ for the $\lambda$ exponent label two sets of solutions which combine to give the full eigenspectrum. The two choices  define the two sectors mentioned above. The quasi-energies are related to $k$ by
\begin{equation}
	\label{eq:eps_k}
\epsilon_j=\frac{\sqrt{(1+\lambda)^2-4\lambda\sin^2 k_j}}{1+\lambda}.
\end{equation}
This quasi-momentum description is convenient for finding EPs which occur at degenerate solutions of Eq.~(\ref{eq:k_cond}), as explored in Section~\ref{sec:eps}.

\begin{figure}[ht]
\includegraphics[width=1.0\linewidth]{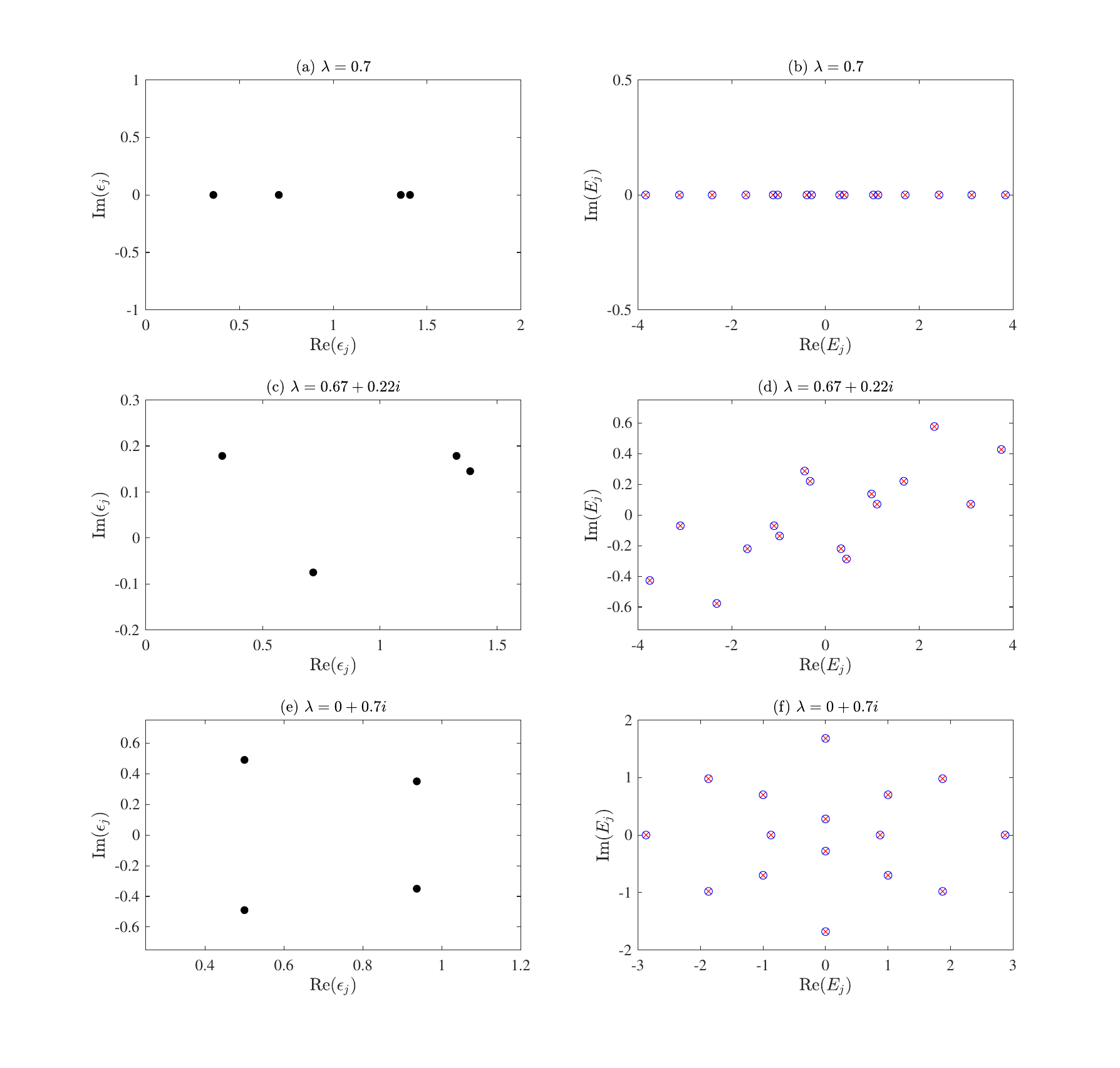}  
\caption{Quasi-energies (left) and corresponding energy spectra (right) for the non-Hermitian XY model with $L=4$, and different values of the complex parameter $\lambda$. The energies (right) show the 16 values obtained via eq.~(\ref{eq:ff}) from the four quasi-energies (blue circles), compared to the values obtained from exact diagonalisation of the full Hamiltonian (red crosses).}
	\label{fig:cl_test}
\end{figure}

\section{Quasi-energy degeneracies and EPs}
\label{sec:eps}

For real $\lambda$, Eq.~(\ref{eq:k_cond}) has $L$ distinct solutions. Lieb et al.'s analysis extends directly to complex values of $\lambda$. 
Figure~{\ref{fig:cl_test}} provides a demonstration of how the free fermion solution works in terms of the quasi-energies for a small test system. For complex $\lambda$, the Hamiltonian is non-Hermitian. 
In particular, if $\lambda$ is allowed to be complex it becomes possible for (\ref{eq:k_cond}) to have a repeated root, resulting in two equal quasi-energies. Following~\cite{henry2023exceptional}, this can be examined analytically by setting both (\ref{eq:k_cond}) and its derivative to be zero, for this model resulting in two equations:
\begin{eqnarray}
\label{eq:kj_ep}
	\sin{(L+2)k} + \lambda^{\pm 1} \sin L k &=& 0 \,,\\ 
	\label{eq:kj_ep2}
	(L+2) \cos{(L+2)k} + L \, \lambda^{\pm 1} \cos L k &=& 0\,.
\end{eqnarray}
These can be combined to eliminate $\lambda$, giving an equation satisfied by special values of the quasimomentum $k_{EP}$:
\begin{equation}
	\label{eq:kj_ep4}
	(L+2)\sin Lk_{EP} \, \cos (L+2)k_{EP} - L\sin (L+2)k_{EP} \, \cos Lk_{EP} = 0 \,.
\end{equation}

A quasi-energy $k_{EP}$ satisfying this equation implies an EP at two corresponding values $\lambda_{EP}$, given by
\begin{equation}
	\label{eq:kj_ep3}
	\lambda_{EP}^{\pm 1} = - \dfrac{\sin(L+2) k_{EP}}{\sin L  k_{EP}}.
\end{equation}
The two choices ($\pm 1$) correspond to EPs in each of the two branches of the spectrum. A single value $k_{EP}$ which solves (\ref{eq:kj_ep4}) corresponds to one EP in each band. Figure~\ref{fig:ep_test_large_xy} shows the values $k_{EP}$ in the complex plane for increasing system sizes, and the corresponding values $\lambda_{EP}$.
Here we have excluded the two trivial solutions $k=0$ and $k=\pi/2$, which respectively correspond to the values $\lambda^{\pm 1}=-(L+2)/L$ and $\lambda^{\pm 1}=(L+2)/L$. 
In fact, these points do exhibit quasi-energy degeneracies, but between quasi-energies in the two different sectors of the Hamiltonian. Their eigenvectors remain distinct, so these points are not EPs, as to be expected since the model is Hermitian for real $\lambda$. 

As an EP example, consider $L=4$. The EPs occur at $\lambda_{EP} = \pm 0.5 \,\mathrm{i}$ and $\lambda_{EP} = \pm 2 \,\mathrm{i}$. 
At $\lambda_{EP} = 2 \,\mathrm{i}$, the four observed EPs have energy eigenvalues $\pm 1$ and $\pm \sqrt{15} \, \mathrm{i}$.
For small system sizes, we have checked numerically that the expected condition $\bra{\mathcal{L}}\ket{\mathcal{R}} =0$ holds exactly at the EPs, where $\bra{\mathcal{L}}$ and $\ket{\mathcal{R}}$ are the relevant left and right eigenvectors of the non-Hermitian Hamiltonian. 
For this model it is possible to undertake an explicit study of the eigenvectors in the vicinity of the EPs making use of the exact formulation of the eigenvectors, which we leave for future work.    

We observe that in total the EPs are located at $2L-4$ points in the complex $\lambda$-plane, appearing in two concentric rings. This is in contrast to the non-Hermitian extension of the quantum Ising chain, for which the EPs are located at $2L$ points in a single ring~\cite{henry2023exceptional}. 

\section{Infinite size limit}

The behaviour of the EPs in the infinite size limit ($L\to\infty$) can be examined by expanding $k_j$ 
following Lieb et al.'s analysis, 
\begin{equation}
	\label{eq:k_exp_L}
	k_j = \frac{\pi j}{L} - \frac{\pi a}{L} + \mathcal{O}\left(\dfrac{1}{L^2}\right),
\end{equation}
where $j \in \{1, 2, \ldots, L\}$. Note that $j$ is of order $L$, so $j/L$ is of order 1. The parameter $a$ is determined by inserting this expansion into (\ref{eq:kj_ep}).
Choosing the $\lambda^{+1}$ branch, this is done by writing
\begin{eqnarray}
	\label{eq:3}
	0 &\approx& \lambda \sin(\pi j -\pi a) + \sin \left(\pi j - \pi a + \dfrac{2\pi j}{L} - \dfrac{2\pi a}{L}\right)\\
	& \approx & \lambda \sin \pi a + \sin \pi a \, \cos\left(\dfrac{2\pi j}{L}\right) + \cos\pi a \,\sin\left(\dfrac{2\pi j}{L}\right) \,.
\end{eqnarray}
This last equation can be rearranged as
\begin{equation}
	\cot{\pi a}  = \dfrac{\lambda + \cos (2\pi j /L)}{\sin(2\pi j /L)} \,.
\end{equation}
The EPs are given by values of $\lambda_{EP}$ and $k_{EP}$ which satisfy both (\ref{eq:kj_ep}) and its derivative (\ref{eq:kj_ep2}), with the latter giving a second equation for $a$ at the EPs:
\begin{equation}
	\label{eq:3a}
	\tan \pi a = - \dfrac{ \frac{L}{L+2}\lambda + \cos{(2\pi j / L)}}{\sin{(2\pi j / L)}}.
\end{equation}
Eliminating $a$ from these last two equations and setting $L / (L+2) \approx 1$ gives
\begin{equation}
	\label{eq:4}
	\dfrac{\sin{(2\pi j/L})}{\lambda+\cos{(2\pi j/L)} } = - \,\dfrac{\lambda + \cos{(2\pi j / L)} }{\sin{(2\pi j / L)}},
\end{equation}
\noindent which reduces to
\begin{equation}
	\label{eq:5}
	\lambda^2 + 2 \lambda\cos{\left(\frac{2\pi j}{L}\right)} + 1 = 0 \, ,
\end{equation}
with solutions
\begin{equation}
	\label{eq:6}
	\lambda = -\cos{\left(\frac{2\pi j}{L}\right)} \pm \mathrm{i} \sin{\left(\frac{2\pi j}{L}\right)} \,.
\end{equation}
Recalling that $j \in \{1,\ldots,L\}$ these are exactly the $L$th roots of unity. However, the solutions $\lambda = \pm 1$ correspond to $k=0$ and $k=\pi/2$ discussed in the previous section. 
Thus for large $L$ the set of EPs corresponds to the $L$ roots of unity, excluding $-1$ and $1$.

\begin{figure}[ht]
    \centering
    \includegraphics[width=0.99\linewidth]{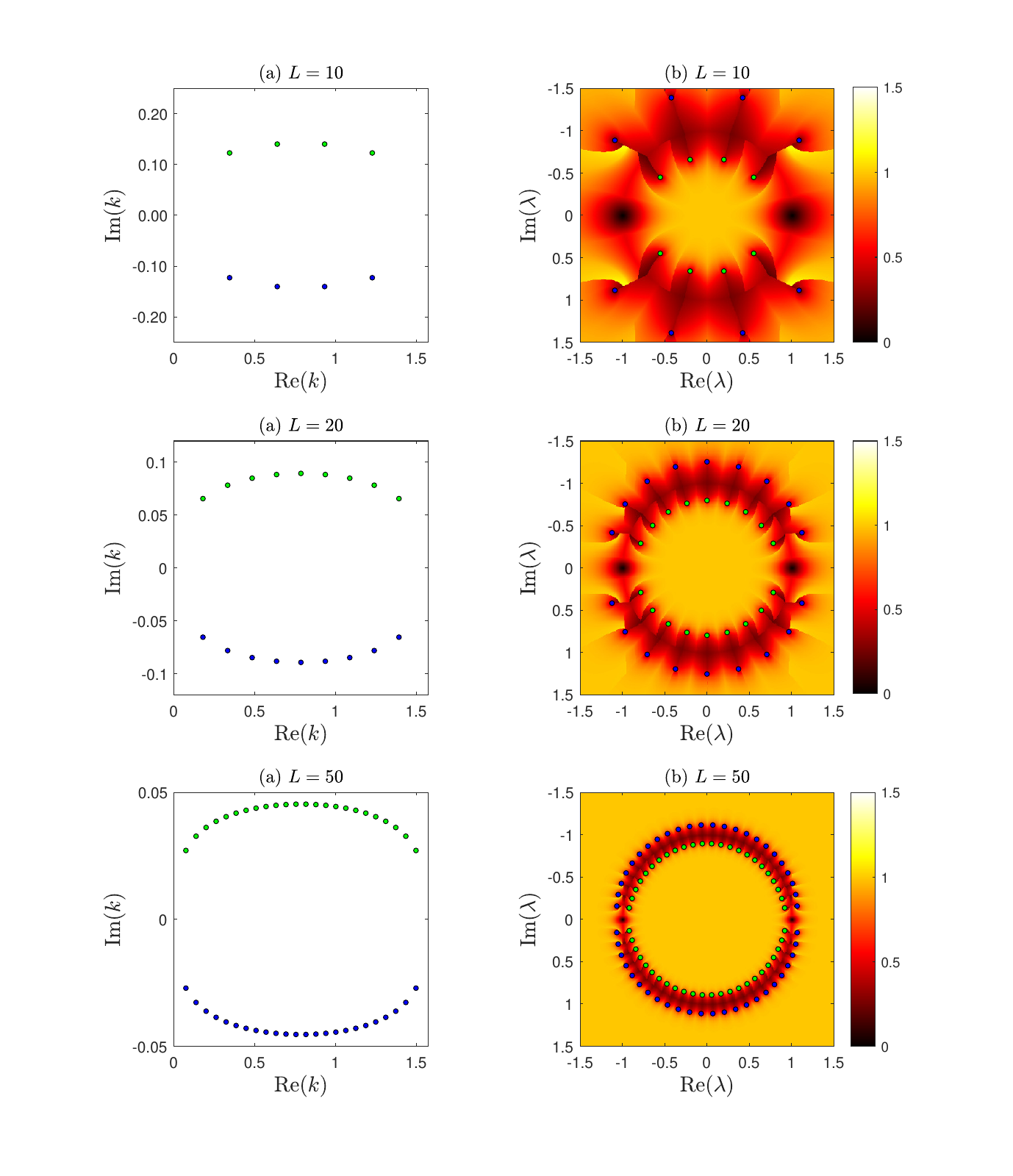}  
    \caption{(left) Solutions of the quasi-energy degeneracy condition (\ref{eq:kj_ep4}) in the complex $k$-plane for increasing values of $L$. 
    Each solution corresponds to an EP (right), where the absolute distance $\Delta\epsilon_{12}$ between the smallest two quasi-energies is shown for comparison. The green and blue circles label degeneracies and thus EPs appearing in the two different rings. }
    \label{fig:ep_test_large_xy}
\end{figure}

The alternative choice of $\lambda^{-1}$ gives the same roots of unity although in a different order, so each EP appears once for $\lambda$ and once for $\lambda^{-1}$, at the same root of unity in the limit $L\to\infty$.
In fact, the next-order corrections for the two cases are different, with each EP ring moving in opposite directions towards the unit circle. This behaviour can be seen in Figure~\ref{fig:ep_test_large_xy}, where the EP rings approach the unit circle from inside and outside as the system size increases.
Another model where EPs can be examined in the limit of large system size is the Lipkin model~\cite{Heiss2005}.


\section{$\PT$-symmetric points}

For the XY model parity can be implemented as reversing the lattice, and time as complex conjugating all operators and coefficients, i.e.,
\begin{eqnarray}
  \label{eq:10}
  \Psym \sigma^x_{j} \Psym &=& \sigma^x_{L-j+1}, \qquad \Tsym \sigma^x_{j} \Tsym = \sigma^x_j, \\
  \Psym \sigma^y_{j} \Psym &=& \sigma^y_{L-j+1}, \qquad  \Tsym \sigma^y_{j} \Tsym = \mathrm{i} \,\sigma^y_j \,.
\end{eqnarray}
It follows that 
\begin{equation}
  \label{eq:11}
  (\PT) \lambda \, \sigma_j^y\sigma^y_{j+1} (\Psym\Tsym)^{-1} = -\bar\lambda \, \sigma_j^y\sigma^y_{j+1} \,.
\end{equation}

The Hamiltonian $H_\lambda$ is thus $\PT$-symmetric if $\lambda = \mathrm{i} \, \lambda_I$, i.e., for pure imaginary coupling.
This can be seen in the bottom row of Figure~\ref{fig:cl_test}, where the quasi-energies and spectrum appear in conjugate pairs and the complex energy spectrum has a reflection symmetry in the real axis, a hallmark of (broken) $\PT$ symmetry.
An interesting related aspect is that $\lambda_{EP}$ can take values exactly on the imaginary $\lambda$ axis if the system size $L$ is a multiple of 4, as can be seen, e.g., in the middle row of 
Figure~\ref{fig:ep_test_large_xy}.
In this way we have identified a subset of four 
$\lambda_{EP}$ values at which the XY model is broken $\PT$-symmetric.

\section{Concluding remarks}

\begin{figure}[t]
\centerline{\includegraphics[width=0.7\linewidth]{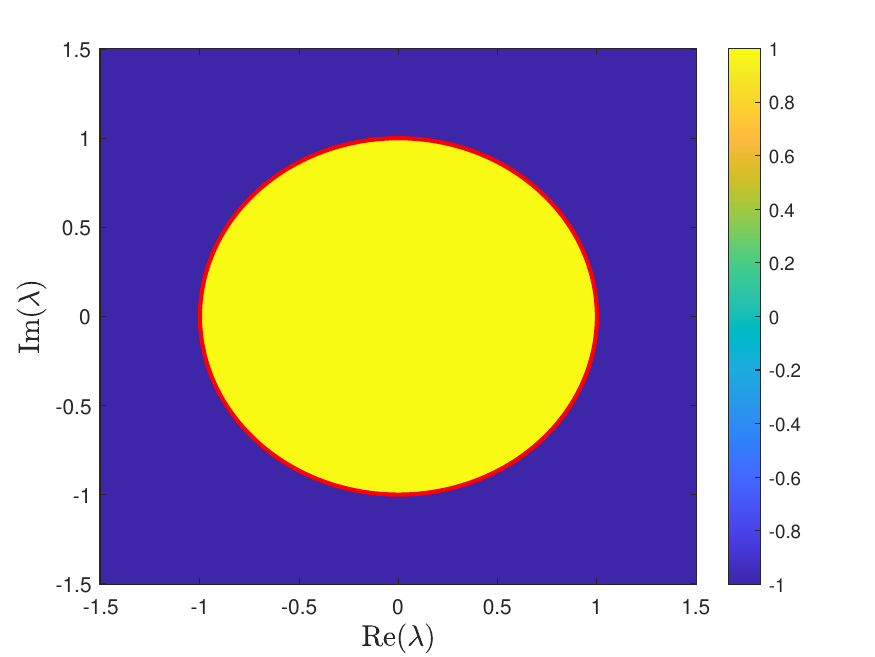}}
	\caption{The topological phase diagram of the non-Hermitian XY model (\ref{eq:1}). The red line is a guide for the eye as the unit circle separating the regions with distinct topological phases $w=1$ and $w=-1$. In the infinite size limit there is a ring of EPs on the unit circle.}
	\label{fig:phase}
\end{figure}

In this work we have identified the location of EPs in the non-Hermitian XY model by taking advantage of the fundamental free fermion structure of the eigenspectrum with open boundary conditions.
The procedure for constructing the eigenspectrum via the quasi-energies is demonstrated in the three indicative cases shown in Figure~\ref{fig:cl_test}. 
%
%
In particular, two quasi-energies may become degenerate for complex parameter $\lambda$, resulting in EPs in the energy eigenspectrum of the model Hamiltonian. 
For finite system size $L$, we have identified two concentric EP rings, as shown in Figure~\ref{fig:ep_test_large_xy}. 
These EP rings converge to the unit circle in the complex $\lambda$-plane as $L$ increases.

This unit circle coincides with the phase boundary characterized by the topological invariants, i.e., by the winding number, demonstrated by rewriting the Hamiltonian in the momentum space~\cite{PhysRevA.97.052115, Liu2025}.
The winding number is simply a measure of winding around the origin, being counted as  
positive (negative) if the curve in the auxiliary plane is traversed along the anticlockwise (clockwise) direction~\cite{XMYang}. 
%
%
Figure~\ref{fig:phase} shows the phase diagram of the non-Hermitian XY model obtained from the non-Hermitian winding number $w$. 
The boundary between the distinct topological phases with $w=1$ and $w=-1$ is the unit circle $\Re(\lambda) ^2+ \Im(\lambda) ^2 =1$. 
We thus see that this phase boundary coincides with the location of the EP rings in the limit $L \to \infty$.
Under the transformation (\ref{trans}) the unit circle in the $\lambda$-plane corresponds to the imaginary $\gamma$-axis in Hamiltonian (\ref{eq:1-1}).  
Conversely, the line of broken $\PT$ symmetry on the imaginary $\lambda$ axis corresponds to the unit circle in the complex $\gamma$-plane.
According to our results, there are four EP locations on this circle of broken $\PT$ symmetry in the complex $\gamma$-plane if the system size is a multiple of 4.

In this study we have considered the location of EPs under open boundary conditions, leaving a possible discussion of periodic boundary conditions for future consideration. 
As recently remarked~\cite{Sirker2024}, beyond the loci of EPs located in the $N$-state free parafermion model~\cite{henry2023exceptional} and the asymmetric hopping model~\cite{Sirker2024}, to which we have added the present study of the non-Hermitian XY model, it will be of interest to consider similar phenomena in other quantum many-body models of this kind, 
such as for example, the class of models discussed in Ref.~\cite{Alcaraz2021}.
Another avenue of potential interest is the possibility of locating Dirac Exceptional Points.
Dirac EPs, having also been observed experimentally~\cite{DiracEP2025}, are a combination of Dirac points typical of Hermitian systems and EPs pertaining to non-Hermitian systems.


\bmhead{Acknowledgements}
This work has been supported by the Australian Research Council's Discovery Program through grant numbers DP210102243 and DP240100838.







\end{document}